\documentclass[intlimits,twoside,a4paper]{article}
\usepackage[cp1251]{inputenc}

\usepackage{subfig}
\usepackage[eqsecnum]{cmpj3}

\usepackage{bm}

\issue{2023}{26}{1}{13201}
\doinumber{10.5488/CMP.26.13201}

\title[Power-law correlated disordered 3D Ising model]%
{Temperature scaling analysis of the 3D disordered Ising model with
power-law correlated defects%
}
\author[S. Kazmin, W. Janke]{
S. Kazmin\orcid{0000-0003-2527-8001}\refaddr{label1,label2}\thanks{\email{stanislav.kazmin@dbfz.de}.},
W. Janke\orcid{0000-0002-5165-9097}\refaddr{label1}\thanks{Corresponding author: \email{wolfhard.janke@itp.uni-leipzig.de}.}
}
\addresses{
\addr{label1} Institut f\"ur Theoretische Physik, Universit\"at Leipzig, IPF 231101,
04801 Leipzig, Germany
\addr{label2} Deutsches Biomasseforschungszentrum gemeinn\"utzige GmbH, 
Torgauer Str.\ 116, 04347 Leizig, Germany
}


\date{Received November 12, 2022, in final form January 17, 2023}

\begin{document}

\maketitle


\begin{abstract}
We consider the three-dimensional site-diluted Ising model with
power-law correlated defects and study the critical behaviour of
the second-moment correlation length and the magnetic susceptibility 
in the high-temperature phase. By comparing, for various defect correlation 
strengths, the extracted critical exponents $\nu$ and $\gamma$
with the results of our previous finite-size scaling study, we 
consolidate the exponent estimates.
\printkeywords
%
\end{abstract}


\section{Introduction}
\label{sec:intro}

It is well known that under certain conditions quenched disorder can affect
the critical behaviour of a physical system. Most extensively 
studied is {\em uncorrelated\/} disorder for which the Harris criterion
\cite{harris} states that 
impurities are irrelevant when the 
specific-heat exponent $\alpha_{\rm pure}$ of the pure system is negative,
whereas for 
$\alpha_{\rm pure} > 0$ 
renormalization-group arguments 
suggest a modified critical behaviour. This prediction has been confirmed in
numerous studies of different models and especially also for the
three-dimensional Ising model 
\cite{ballesteros1998,calabrese,berche_cpc02,berche_munich03,berche_nic03,berche_athens04,berche_epjb04,berche_cmp05,berche_lat2005,murtazaev,hasenbusch2007}
for which $\alpha_{\rm pure} \approx 0.1102$ \cite{cb,ferrenberg_xu_landau}.

In realistic physical systems, however, it is more likely that the impurities 
or defects exhibit some kind of spatial correlations. When these correlations 
decay sufficiently slowly, e.g., they follow asymptotically, for large distances $r$, 
the power law $r^{-a}$ with a correlation exponent 
$a < d$, where $d$ is the dimension of the system, one observes 
a new scenario for long-range {\em correlated\/} (quenched) disorder: 
An extension of the Harris criterion by Weinrib and Halperin \cite{weinrib} 
and the later considerations \cite{nalimov,korzhenevskii1,korzhenevskii2} predict 
that in this case the correlation-length exponent $\nu$ obeys quite generally
\begin{equation}
\nu = 2/a \,.
\label{eq:WH-nu}
\end{equation}
Similar predictions for other critical exponents read
$\alpha = 2 - d\nu = 2(a-d)/a$,
$\beta = (2-\epsilon)/a + {\cal O}(\epsilon^2)$, and
$\gamma = 4/a + {\cal O}(\epsilon^2)$,
where $\epsilon = 4-d$ together with 
$\delta = 4-a$ enters the employed
$\epsilon$-$\delta$ renormalization-group expansion \cite{weinrib}. As already speculated
in  \cite {weinrib}, the relation (\ref{eq:WH-nu}) plays a
special role and is expected to
be valid to all orders of this expansion \cite{nalimov}.
%
%
%
More recently this relation was confirmed for the special case of the 
two-dimensional Ising model via a mapping to Dirac fermions and applying an
alternative renormalization-group scheme with a double expansion in $\epsilon' = 2-d$
and $\delta' = 2-a$ up to two-loop order, $\nu = 2/a + O(\delta'^3)$ 
\cite{dudka2016}. 
On the other hand, when the correlations decay more rapidly, e.g., 
exponentially or power-law-like with $a \ge d$, one falls back 
into the universality class of uncorrelated disorder.

In  \cite{own_prb20,own_prb22}, we  studied the power-law
correlated case for the site-diluted three-dimensional (3D) Ising model
with extensive Monte Carlo (MC) computer simulations in the vicinity
of criticality by employing finite-size scaling (FSS) techniques for
the data analyses. Here, we complement these studies by reporting alternative 
estimates for the critical exponents $\nu$ and $\gamma$ obtained from the
analyses of the temperature scaling 
of MC data for the second-moment correlation 
length and magnetic susceptibility when approaching the critical
point in the high-temperature phase.

The rest of the paper is organized as follows. In section\ \ref{sec:model}
we briefly recall the employed model and the simulation method. Our
results are presented and discussed in section\ \ref{sec:results},
and in section\ \ref{sec:concl} we conlude the paper with a summary and 
brief outlook to future work.


\section{Model and methods}
\label{sec:model}

The three-dimensional Ising model with site disorder is defined by the
Hamiltonian
\begin{equation}
{\cal H} = -J \sum_{\langle ij \rangle} \epsilon_i \epsilon_j s_i s_j,
\label{eq:H}
\end{equation}
where the spins $s_i$ take on the values $\pm 1$ and the sum
runs over all nearest-neighbor pairs denoted by $\langle ij \rangle$
of a simple-cubic lattice of size $V = L^3$ with periodic
boundary conditions. The defect variables are $\eta_i = 1 - \epsilon_i = 0$ when a
spin is present at site $i$ and $\eta_i = 1$ when site $i$ is empty, 
i.e., occupied by a defect.
The coupling constant is set to $J = 1$, fixing the unit of energy and,
by setting the Boltzmann constant $k_{\text{B}} = 1$, also the temperature scale.

For uncorrelated disorder, the defects are chosen randomly according to
the probability density
\begin{equation}
f(\eta) = p \delta_{\eta,0} + p_d \delta_{\eta,1},
\label{eq:eta_distr}
\end{equation}
where $\delta_{i,j}$ is the Kronecker delta symbol.
Here, $p_d$ denotes the concentration of defects and $p = 1 - p_d$ is 
the concentration of spins.\footnote{Note that in the corresponding definition
in  \cite{own_prb20,own_prb22}, $p$ and $p_d$ are inadvertently interchanged.}
We use the grand-canonical approach where the desired defect concentration $p_d$ 
is the mean value over all the considered disorder realizations.

For correlated disorder, we  additionally introduce a long-range spatial correlation 
between the defects at sites $i$ and $j$ that decays asymptotically for 
large distances $r_{ij}$ according to the power law,
\begin{equation}
\langle \eta_i \eta_j \rangle_c \equiv 
\langle (\eta_i - \langle \eta_i \rangle)(\eta_j - \langle \eta_j \rangle) \rangle 
\propto \frac{1}{r_{ij}^a} \,,
\label{eq:correlation}
\end{equation}
where $a>0$ is the correlation exponent. For the numerical generation 
of the defect correlation we employed the Fourier filter method described 
by Makse  et al. \cite{makse,makse2} in the publicly available 
C++ implementation\footnote{The C++ code
is available at github.com/CQT-Leipzig/correlated\_disorder.} of 
 \cite{own_corr-gen},
which for technical reasons 
considers a slightly modified correlation function $C(r) = (1+r^2 )^{-a/2}$ 
that agrees asymptotically with (\ref{eq:correlation}) (see also
 \cite{own_corr-gen0}). The resulting $1/r^2$ 
corrections in combination with finite-size effects make the measurement of the 
actual correlation exponent $\overline{a} \approx a$ an important analysis step,
cf.\ table \ref{tab:temperature_scaling_analysis:compare_fss_ts}; for details we 
refer to  \cite{own_prb20}. 

We considered the correlation exponents $a = 1.5$, 
$2.0$, $2.5$, $3.0$, $3.5$, and $\infty$ (standing symbolically for the 
uncorrelated case) and studied in each case the eight defect concentrations 
$p_d = 0.05$, $0.1$, $0.15$, $0.2$, $0.25$, $0.3$, $0.35$, and $0.4$. 
For each disorder realization, the MC simulations of this model were 
performed at various temperatures 
$T$ 
with the Swendsen-Wang 
multiple-cluster update algorithm \cite{sw-multiple}, collecting $N = 10\,000$
measurements after 500 thermalization sweeps. All final results are the averages
over $N_c = 1000$ randomly chosen disorder realizations.
The linear lattice 
size was taken for all temperatures $T$ to be $L = 256$, the largest lattice of our FSS 
studies \cite{own_prb20,own_prb22}. Since the correlation length $\xi(T)$ 
and hence finite-size effects quickly diminish away from the critical point,
one could in principle adapt the lattice size to satisfy $L \gg \xi(T)$. 
However, in the case of correlated defects, the measured correlation 
exponent $\overline{a}$ was found to be slightly $L$-dependent \cite{own_prb20} so 
that mixing different lattice sizes in scaling analyses of the high-temperature
data should be avoided.

We studied two observables, the second-moment
correlation length $\xi$ calculated as \cite{wj-greifswald}
\begin{equation}
\xi = \frac{1}{2\sin(\piup/L)}\sqrt{\frac{S(\mathbf{0})}{S(\mathbf{1})} - 1},
\label{eq:xi}
\end{equation}
where $S(\mathbf{k})$ is the discrete Fourier transform of the spatial
spin-spin correlation function $\langle s_i s_j \rangle$ 
in the high-temperature phase
evaluated
at $\mathbf{0} = (0,0,0)$ and $\mathbf{1} = (2\piup/L,0,0)$, and
the (high-temperature) susceptibility 
\begin{equation}
\tilde{\chi} = \beta V \langle m^2 \rangle,
\end{equation}
where $m= (1/V) \sum_i \epsilon_i s_i$ is the magnetization density and
$\beta \equiv 1/T$. Note that
$S(\mathbf{0}) = \tilde{\chi}/\beta$.

When $T$  approaches the critical temperature $T_c$, the expected 
temperature-scaling behaviour of the disorder-averaged observables (indicated 
by $[\dots]$) reads
\begin{eqnarray}
[\xi(T)] &=& a |t|^{-\nu} (1 + \dots)\,,          \\
\label{eq:xi_scal}
[\tilde{\chi}(T)] &=& b |t|^{-\gamma} (1 + \dots),
\label{eq:chi_scal}
\end{eqnarray}
where $t = (1-T/T_c)$ $(\leqslant 0)$ is the reduced temperature and
$(1 + \dots)$ indicates analytical and confluent scaling corrections
which vanish as $T \to T_c$. 


\section{Results}
\label{sec:results}

Using the sufficiently precise estimates of $T_c$ from  \cite{own_prb22},
we performed linear fits of $\ln [\xi(T)]$ and $\ln [\tilde{\chi}(T)]$ in 
$\ln |t|$ which provided us with the estimates for critical exponents $\nu$ and 
$\gamma$, respectively. In what follows, we describe the analysis steps 
for the observable $\xi$ and the exponent $\nu$ in some detail and then present an
analogous brief discussion for $\tilde{\chi}$ and the exponent $\gamma$ .

\begin{figure}[!t]
\centering
\subfloat[$a = \infty$]{\includegraphics[scale=0.9]{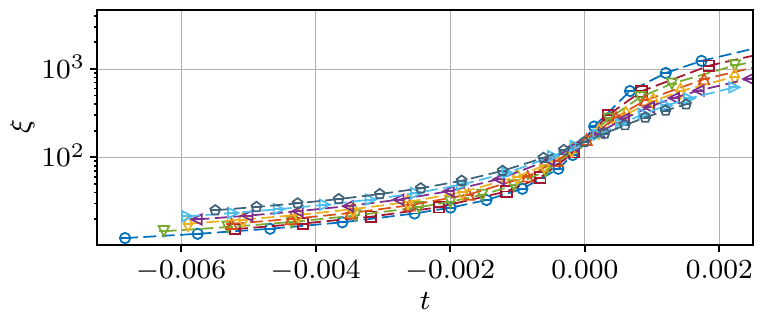}} 
\subfloat[$a = 3.5$]{\includegraphics[scale=0.9]{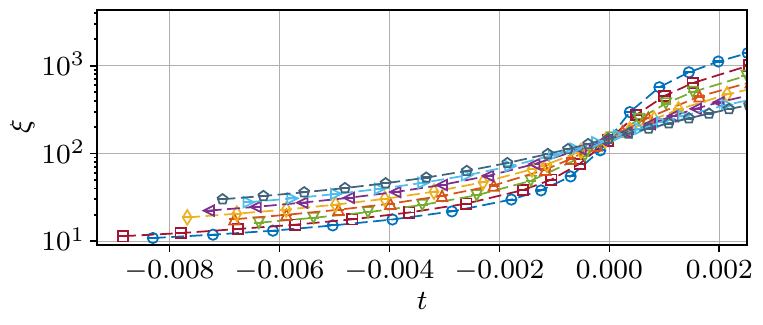}}
\vspace*{-3mm}
\subfloat[$a = 3.0$]{\includegraphics[scale=0.9]{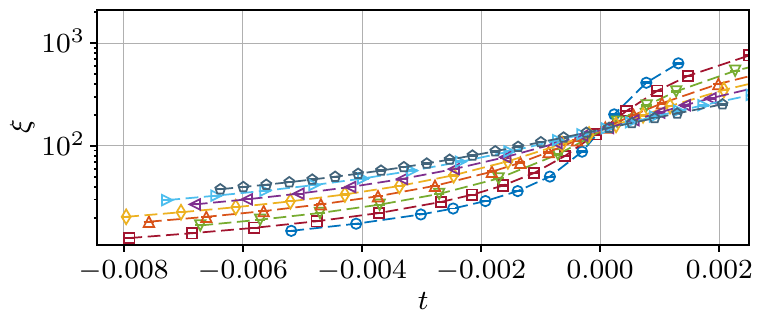}}
\subfloat[$a = 2.5$]{\includegraphics[scale=0.9]{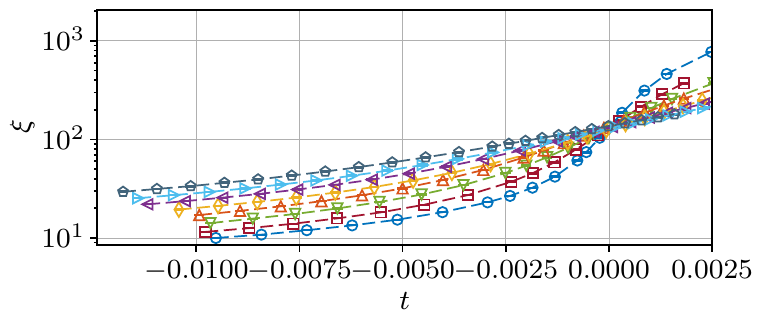}}
\vspace*{-3mm}
\subfloat[$a = 2.0$]{\includegraphics[scale=0.9]{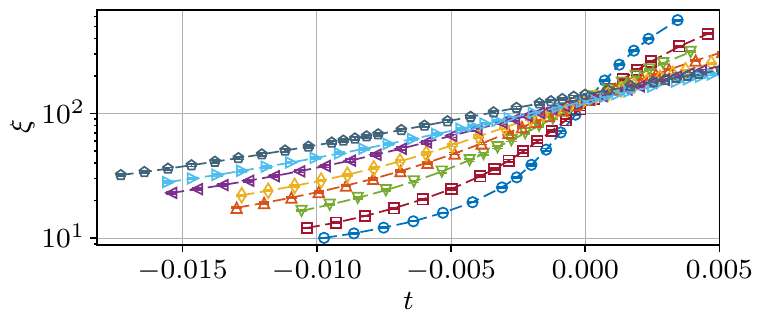}}
\subfloat[$a = 1.5$]{\includegraphics[scale=0.9]{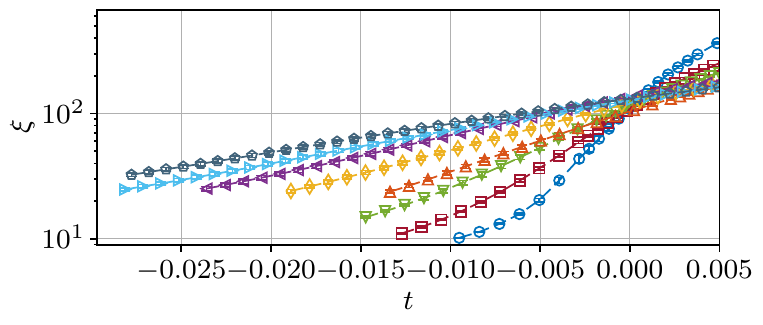}}
\vspace*{-3mm}
\subfloat{\includegraphics[scale=1]{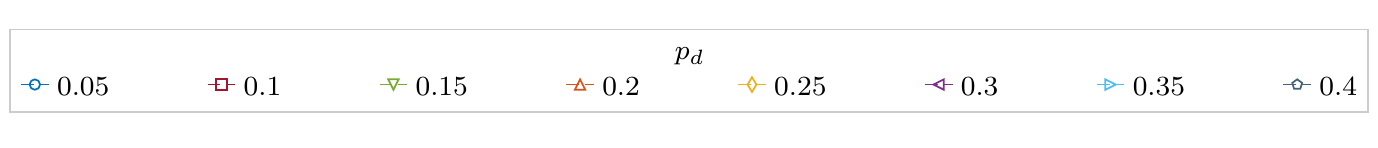}}
\caption
[Correlation length $\xi$ as function of the reduced temperature $t$.]
{(Colour online) Correlation length $[\xi(T)]$ as function of the reduced temperature $t$ for all
considered correlation exponents $a$ and defect concentrations $p_d$. The 
definition (\ref{eq:xi}) of $\xi$ is valid only in the high-temperature phase with 
$t \leqslant 0$, but we extended the curves in order to see the crossing points 
at $t \approx 0$ better.}
\label{fig:temperature_scaling_analysis:xi_exp_t_curves}
\end{figure}


\subsection{Critical exponent \texorpdfstring{$\nu$}{\nu}}

By plotting the disorder averaged correlation length $[\xi(T)]$ for  various defect 
concentrations $p_d$ as a function of $t$, we visually verified the  
used $T_c$ estimates (that depend on both $a$ and $p_d$), as can be seen in figure
\ref{fig:temperature_scaling_analysis:xi_exp_t_curves}
where a negative reduced temperature $t \leqslant 0$ corresponds to the 
high-temperature phase. 
All the curves for different $p_d$  intersect at 
$t \approx 0$. Only for the strongest correlations with $a \leqslant 2.0$, 
visually  not all of them intersect in one point. Strictly speaking, 
$\xi$ is defined only in the high-temperature phase where
$t \leqslant 0$, but we extended it to $t > 0$ in order to see the intersections 
better.

\begin{figure}[h]
\centering
\subfloat[$a = \infty$]{\includegraphics[scale=0.9]{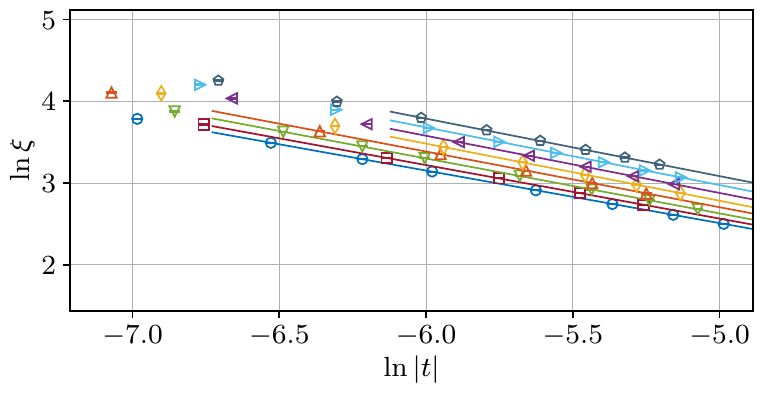}}
\subfloat[$a = 3.5$]{\includegraphics[scale=0.9]{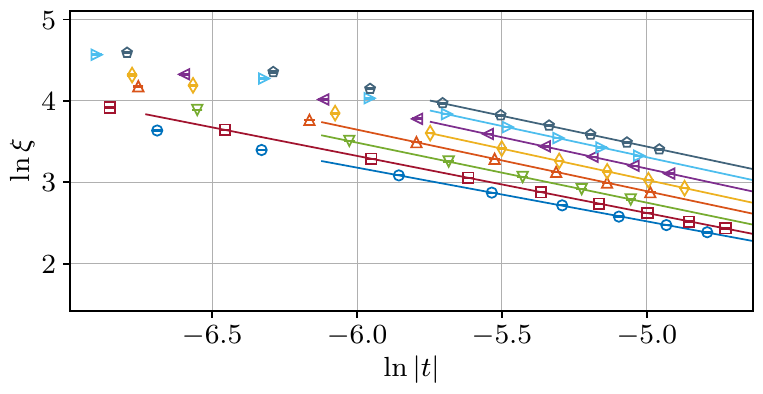}}
\vspace*{-3mm}
\subfloat[$a = 3.0$]{\includegraphics[scale=0.9]{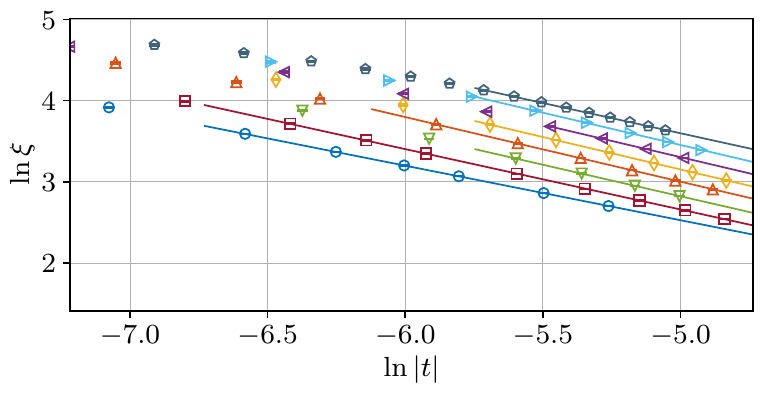}}
\subfloat[$a = 2.5$]{\includegraphics[scale=0.9]{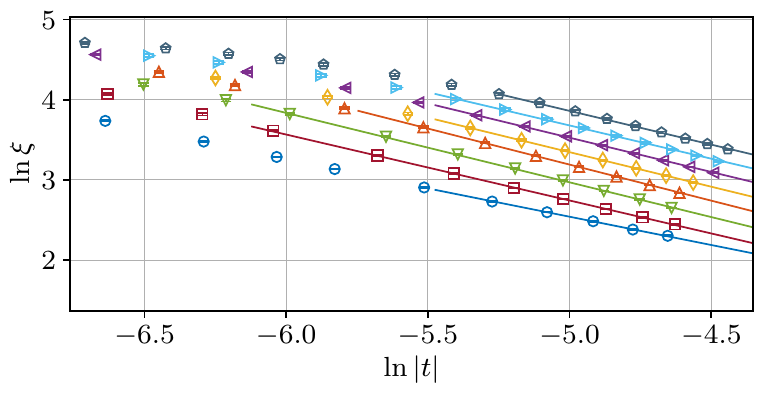}}
\vspace*{-3mm}
\subfloat[$a = 2.0$]{\includegraphics[scale=0.9]{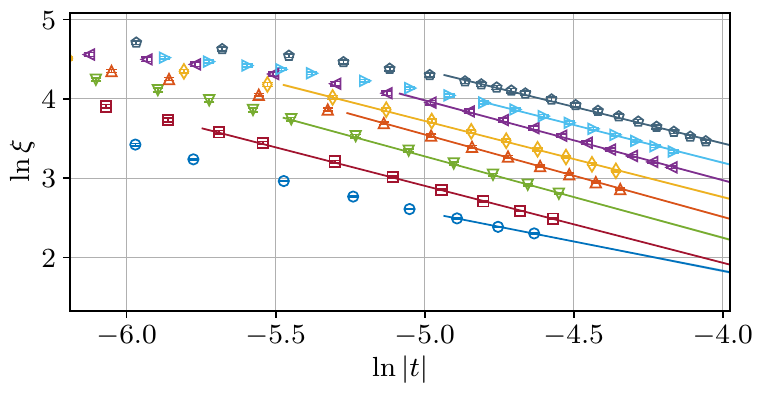}}
\subfloat[$a = 1.5$]{\includegraphics[scale=0.9]{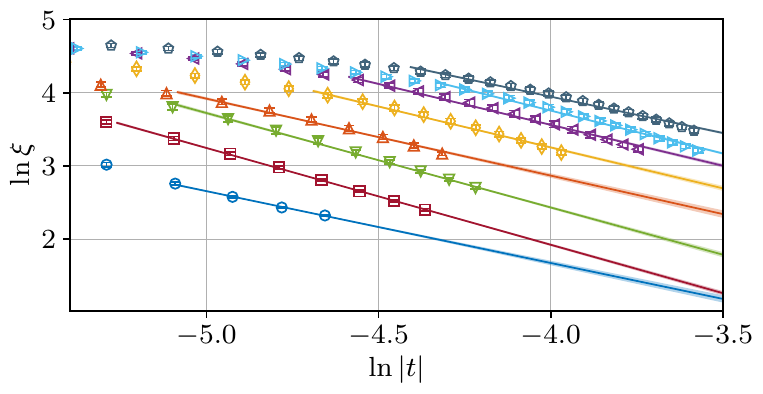}}
\vspace*{-3mm}
\subfloat{\includegraphics[scale=1]{figures/final/xi_scaling/xi_pd_for_legend}}
\caption
[Examples of the fits of $\xi(t)$.]
{(Colour online) Examples of the fits of $[\xi(t)]$ with the ansatz 
(\ref{eq:temperature_scaling_analysis:xi_fit_ansatz}).
For each concentration of defects $p_d$ the smallest $|t|_{\rm min}$ for which 
$\chi^2 \leqslant 1.0$ was true for the first time is used in the plots.}
\label{fig:temperature_scaling_analysis:xi_exp_fit_examples}
\end{figure}

\begin{figure}[h]
\centering
\subfloat[$a = \infty$]{\includegraphics[scale=0.9]{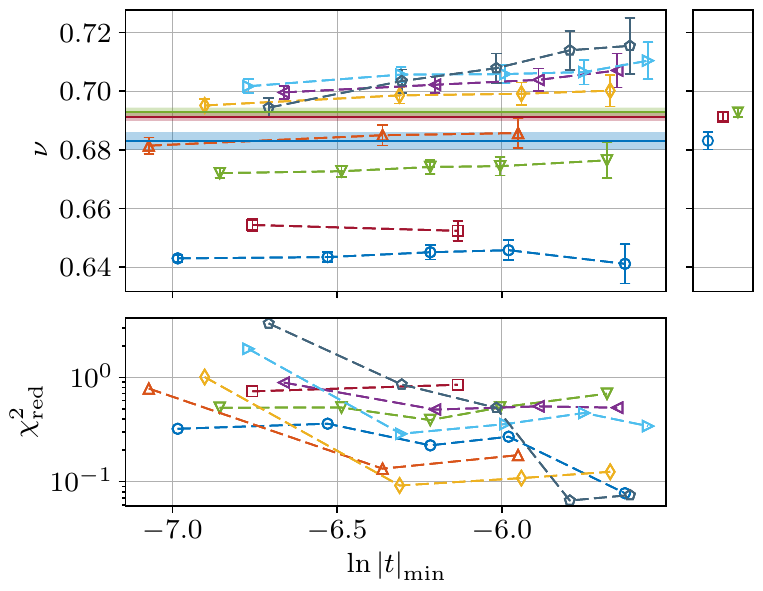}}
\subfloat[$a = 3.5$]{\includegraphics[scale=0.9]{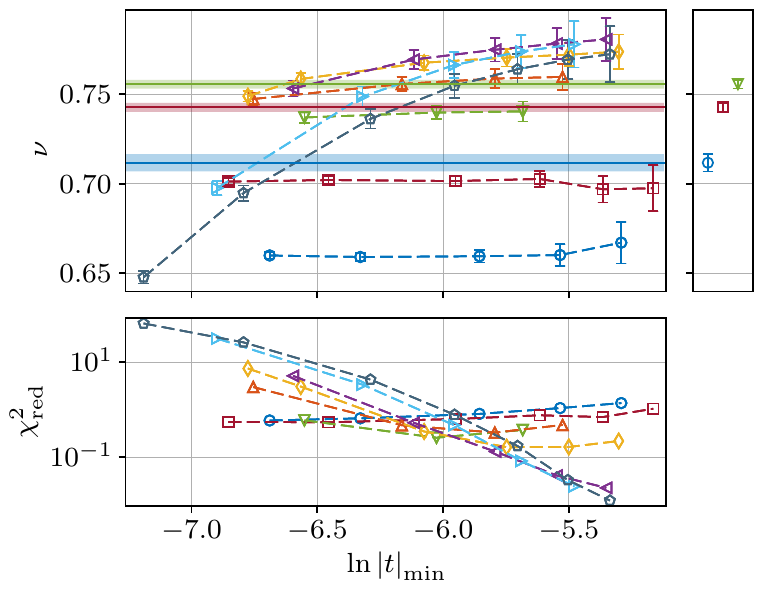}}
\vspace*{-3mm}
\subfloat[$a = 3.0$]{\includegraphics[scale=0.85]{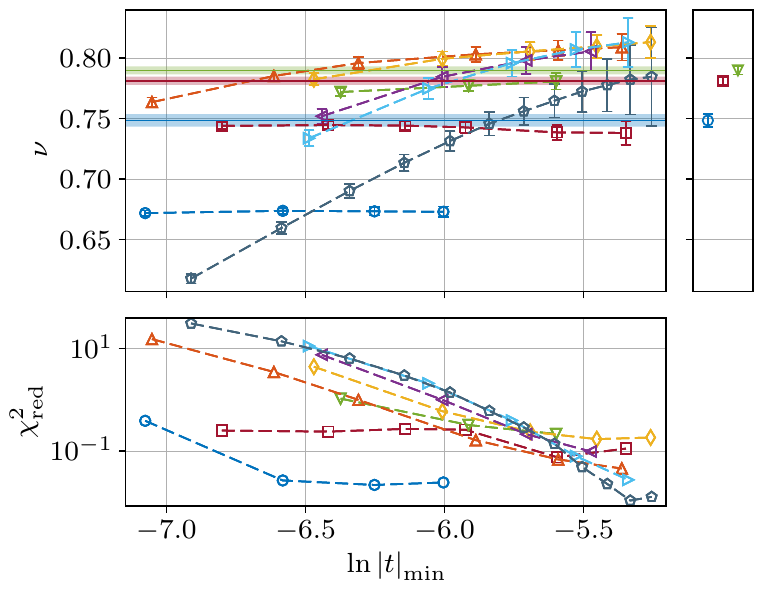}}
\subfloat[$a = 2.5$]{\includegraphics[scale=0.85]{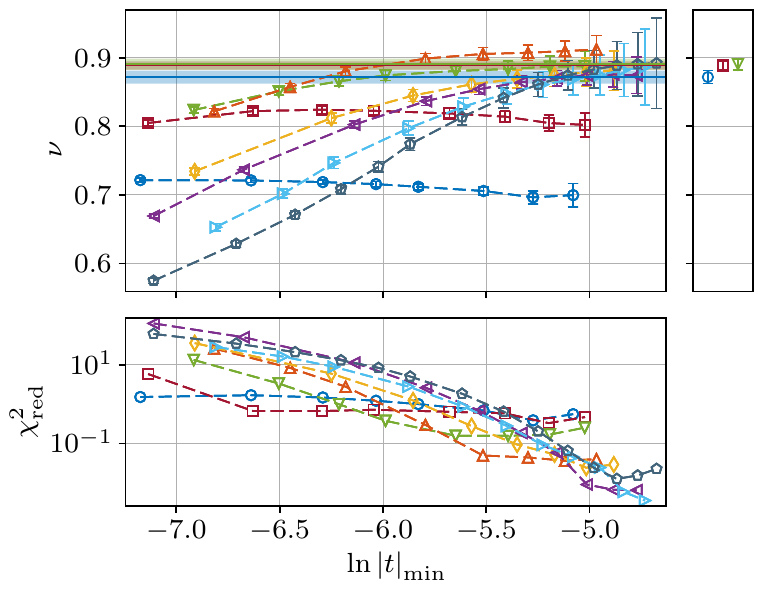}}
\vspace*{-3mm}
\subfloat[$a = 2.0$]{\includegraphics[scale=0.85]{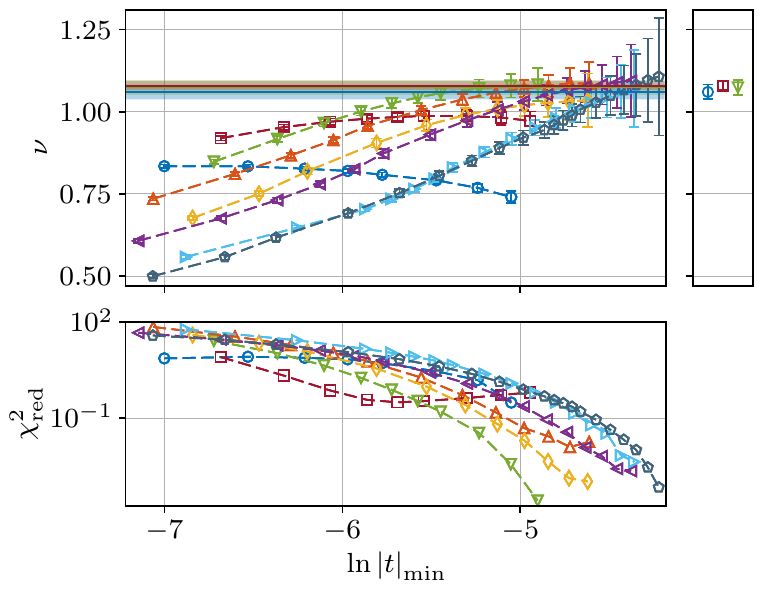}}
\subfloat[$a = 1.5$]{\includegraphics[scale=0.85]{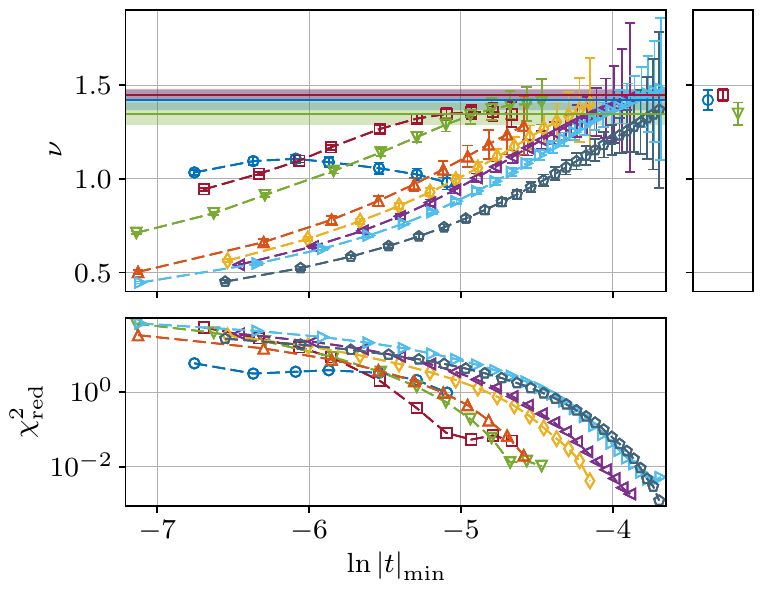}}
\vspace*{-3mm}
\subfloat{\includegraphics[scale=1]{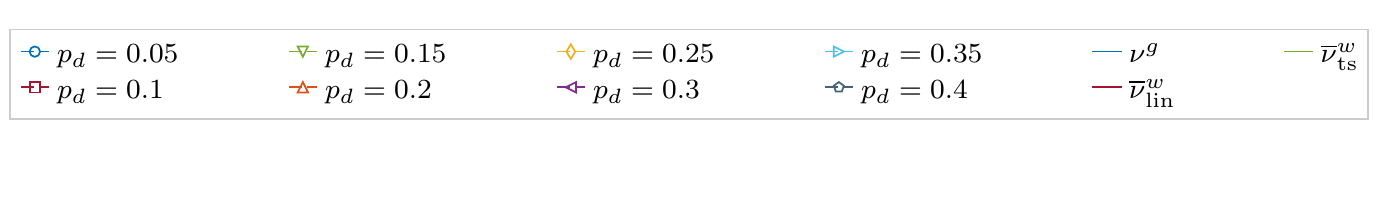}}
\vspace*{-7mm}
\caption
[Final fit results of $\xi(t)$ for different $|t|_{\rm min}$.]
{(Colour online) Final fit results of $[\xi(t)]$ using the ansatz
(\ref{eq:temperature_scaling_analysis:xi_fit_ansatz}) for different 
$|t|_{\rm min}$ and all concentrations of defects $p_d$.
The weighted means $\overline{\nu}^w_{\text{ts}}$ over all concentrations with
$p_d \geqslant 0.15$ and for the largest possible $|t|_{\rm min}$ are shown together 
with the results 
$\overline{\nu}^w_{\rm lin}$ 
and 
$\nu^{\rm g}$ 
from the FSS 
analysis \cite{own_prb22,SK_Thesis}. The narrow right-hand panels show a separate 
comparison between the different estimates for $\nu$ which are plotted as horizontal 
lines in the main plots.}
\label{fig:temperature_scaling_analysis:nu_t_min_dependence}
\end{figure}

Considering only the high-temperature values with $t \leqslant 0$, we performed 
for each correlation exponent $a$ and each defect concentration $p_d$ 
an individual fit with the ansatz
\begin{align}
\ln [\xi(t)]  = A - \nu \ln |t| \,.
\label{eq:temperature_scaling_analysis:xi_fit_ansatz}
\end{align}
Since the power-law behaviour only starts at a certain distance away from 
$t = 0$, i.e., once finite-size effects become neglectable, 
we varied the smallest $|t|_{\rm min}$ included in the fits from its 
smallest value near $t = 0$ to a maximum value where only three degrees of 
freedom remained. Examples of the fits are presented in 
figure \ref{fig:temperature_scaling_analysis:xi_exp_fit_examples}
which shows one main problem with this procedure. We clearly observe 
finite-size effects for each $p_d$, since the data points 
curve down as $|t| \to 0$. Compare, e.g., 
the plot for $a = 1.5$ where this effect is 
most pronounced. However, since the statistical errors 
(estimated with the Jackknife method \cite{jackknife}) 
are quite large, 
linear fits still provide reasonable $\chi^2_{\rm red}$ values
per degree of freedom. The resulting estimates 
of the exponent $\nu$ together with the $\chi^2_{\rm red}$ values 
for each $a$ and $p_d$ and for all the considered 
$|t|_{\rm min}$ are shown in 
figure \ref{fig:temperature_scaling_analysis:nu_t_min_dependence}.
The data show a clear dependence on the defect concentration $p_d$ and also 
on $|t|_{\rm min}$. They do not reach the plateau value even for the largest 
$|t|_{\rm min}$ and the estimates for $p_d \leqslant 0.1$ are clearly influenced by 
the crossover to the pure Ising model. This reflects the observation in 
figure \ref{fig:temperature_scaling_analysis:xi_exp_t_curves} that 
$[\xi]$ exhibits the strongest curvature for the smallest defect concentration
$p_d$. Therefore, for a quantitative comparison, we computed the error weighted 
mean $\overline{\nu}_{\rm ts}^w$ over all estimates for $p_d \geqslant 0.15$ where 
for each $p_d$ we used the fit with the largest possible $|t|_{\rm min}$ having
three degrees of freedom. This may be not an optimal solution but at 
least it was closer to possible plateau values than using the fits 
with the smallest possible $|t|_{\rm min}$ for which $\chi^2_{\rm red} \leqslant 1.0$ was 
satisfied for the first time. The latter results are way too low and clearly 
do not represent the asymptotic behaviour. As mentioned above, this is 
due to the relatively large statistical errors which made the 
simple linear fits acceptable, even though finite-size effects were still 
present. 

The weighted means $\overline{\nu}_{\rm ts}^w$ are compared 
in the narrow right-hand panels of 
figure \ref{fig:temperature_scaling_analysis:nu_t_min_dependence}
with the estimates from our FSS analysis:
Weighted means $\overline{\nu}^w_{\rm lin}$ of individual linear fits
neglecting the scaling corrections \cite{SK_Thesis}
and $\nu^{\rm g}$ from non-linear ``global'' fits 
including corrections-to-scaling \cite{SK_Thesis,own_prb22,yurko_omega}.
The numerical values are compiled in 
table \ref{tab:temperature_scaling_analysis:compare_fss_ts}
where we additionally include
the weighted means $\overline{\nu}^w$
of non-linear FSS fits including the scaling 
corrections \cite{SK_Thesis}.
Except for $a = 1.5$, the estimates 
$\overline{\nu}^w_{\text{ts}}$ 
obtained from temperature scaling
are closer to 
$\overline{\nu}^w_{\rm lin}$ than to 
$\overline{\nu}^w$ respectively $\nu^{\rm g}$ and are slightly larger.
The value for $a = 1.5$ is possibly smaller because the estimates for larger 
$p_d$ show very large errors and hence the smallest included defect concentration 
$p_d = 0.15$ dominates the weighted mean. 
Hence, the estimate for this value of $a$ should  be considered with some reservation 
even though the exemplary fits for $a = 1.5$ displayed in figure 
\ref{fig:temperature_scaling_analysis:xi_exp_fit_examples}(f)
do not look particularly worrying. 
The biggest deviations can be seen for the two correlation exponents 
$a = 3.0$ and $3.5$ which is exactly the same behaviour as for the two types of 
FSS estimates, i.e., $\overline{\nu}^w_{\rm lin}$ and~$\overline{\nu}^w$~or~$\nu^{\rm g}$. We interpret this as a signal for the theoretically expected crossover
from the correlated to effectively uncorrelated behaviour at $a \approx d=3$.

Although one should take the estimates $\overline{\nu}^w_{\text{ts}}$ with 
some care, we nevertheless can qualitatively confirm the FSS results in 
all the considered cases. The prediction (\ref{eq:WH-nu}) of Weinrib and 
Halperin \cite{weinrib} that $\nu = 2/a$ is not matched quantitatively. 
The results lie above this prediction, but the dependence on $a$ 
respectively the measured $\overline{a}$ is clearly in accordance with the 
FSS results which indeed show a $\propto 1/\overline{a}$ behaviour \cite{own_prb22}. 
For a comparison with previous results for selected cases of $a$ and $p_d$
by other groups 
\cite{ballesteros_parisi,prudnikov2000,prudnikov2005,ivaneyko_berche-et-al,wang2019},
we refer to table I and to the discussion in  \cite{own_prb20,own_prb22}.
Let us finally note that we also have checked the influence of the statistical
error of the $T_c$ estimates on the results, but it turned out that it can 
be neglected due to much larger errors coming from the fits themselves.


\subsection{Critical exponent \texorpdfstring{$\gamma$}{\gamma}}

\begin{figure}[!t]
\centering
\subfloat[$a = \infty$]{\includegraphics[scale=0.9]{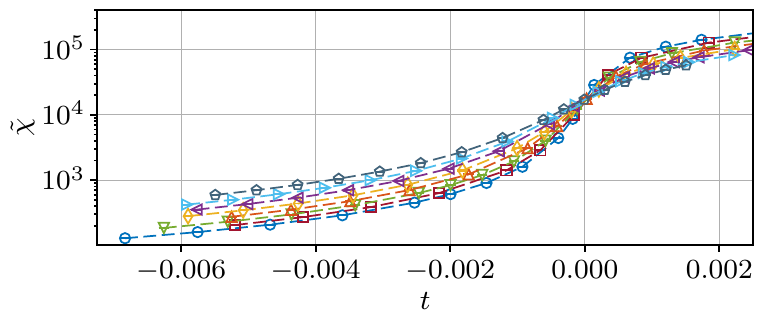}}
\subfloat[$a = 3.5$]{\includegraphics[scale=0.9]{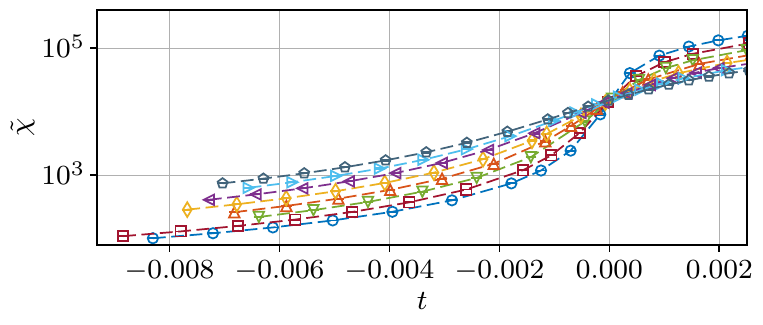}}
\vspace*{-3mm}
\subfloat[$a = 3.0$]{\includegraphics[scale=0.9]{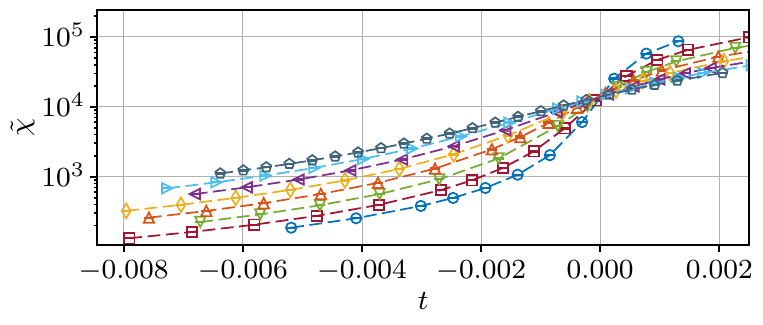}}
\subfloat[$a = 2.5$]{\includegraphics[scale=0.9]{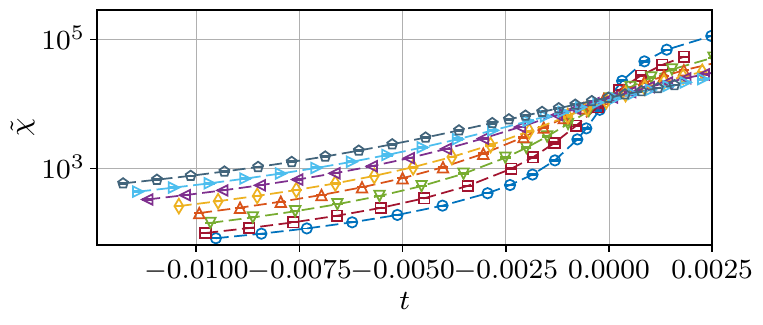}}
\vspace*{-3mm}
\subfloat[$a = 2.0$]{\includegraphics[scale=0.9]{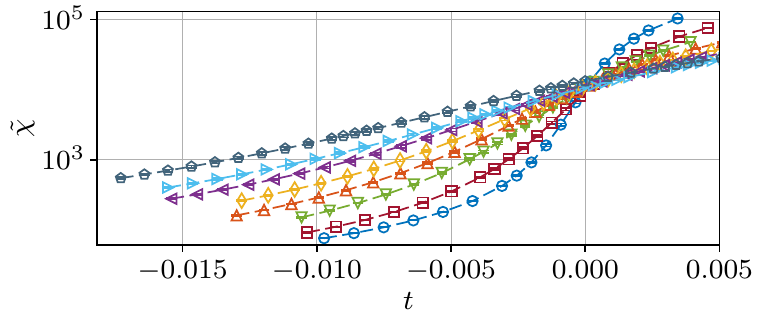}}
\subfloat[$a = 1.5$]{\includegraphics[scale=0.9]{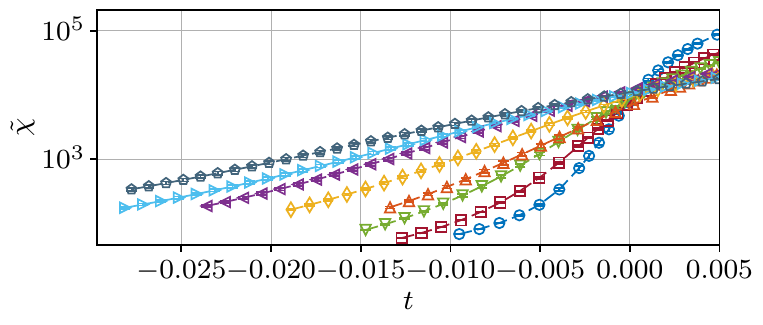}}
\vspace*{-3mm}
\subfloat{\includegraphics[scale=1]{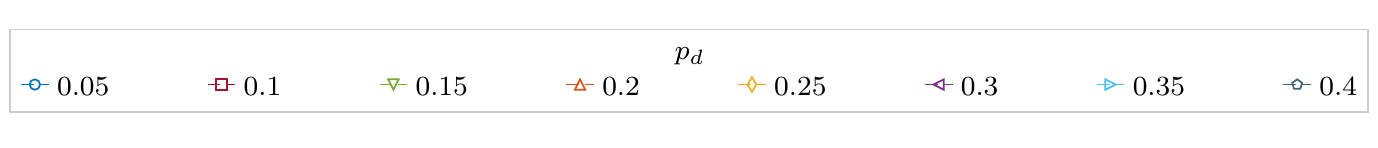}}
\caption
[Susceptibility $\tilde \chi$ as function of the reduced temperature $t$.]
	{(Colour online) Susceptibility $[\tilde \chi (T)]$ as function of the reduced temperature $t$.
The definition of $\tilde \chi$ is valid only in the high-temperature phase 
with $t \leqslant 0$, but we extended the curves in order to see the crossing 
points better.}
\label{fig:temperature_scaling_analysis:chiht_exp_t_curves}
\end{figure}

\begin{figure}[!t]
\centering
\subfloat[$a = \infty$]{\includegraphics[scale=0.9]{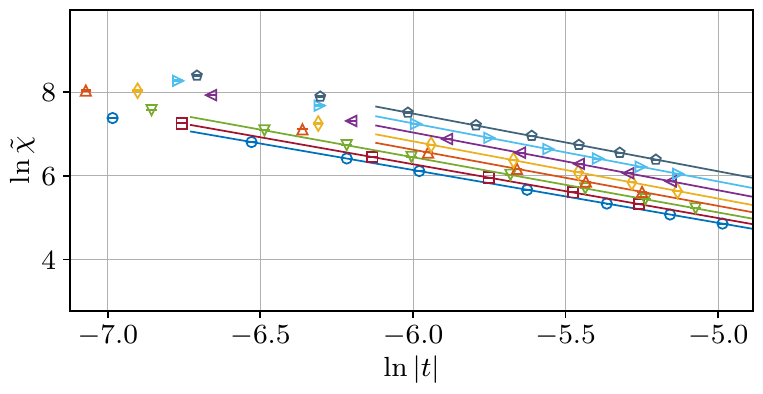}}
\subfloat[$a = 3.5$]{\includegraphics[scale=0.9]{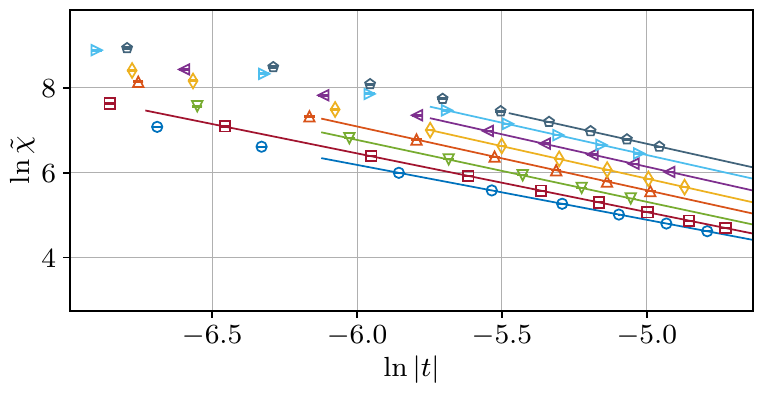}}
\vspace*{-3mm}
\subfloat[$a = 3.0$]{\includegraphics[scale=0.9]{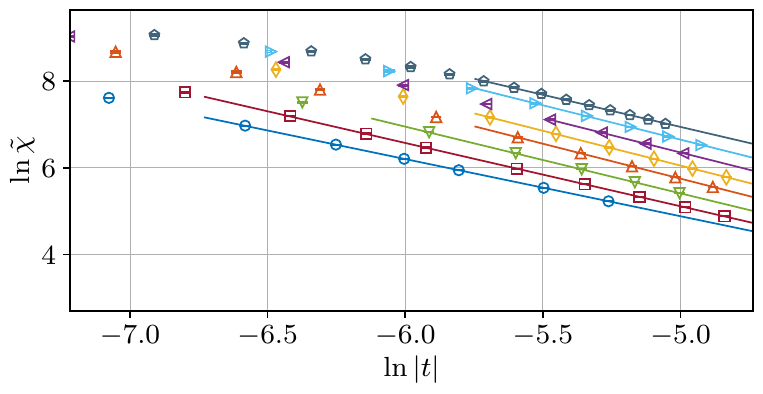}}
\subfloat[$a = 2.5$]{\includegraphics[scale=0.9]{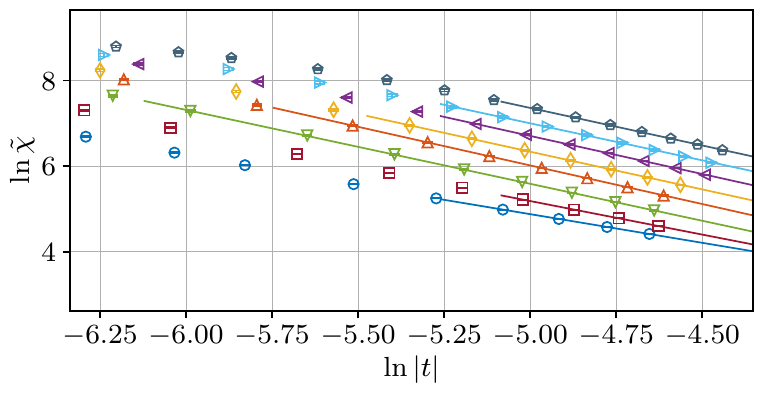}}
\vspace*{-3mm}
\subfloat[$a = 2.0$]{\includegraphics[scale=0.9]{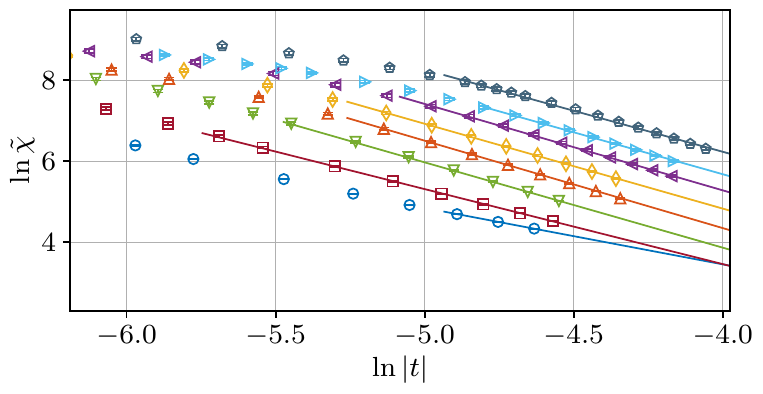}}
\subfloat[$a = 1.5$]{\includegraphics[scale=0.9]{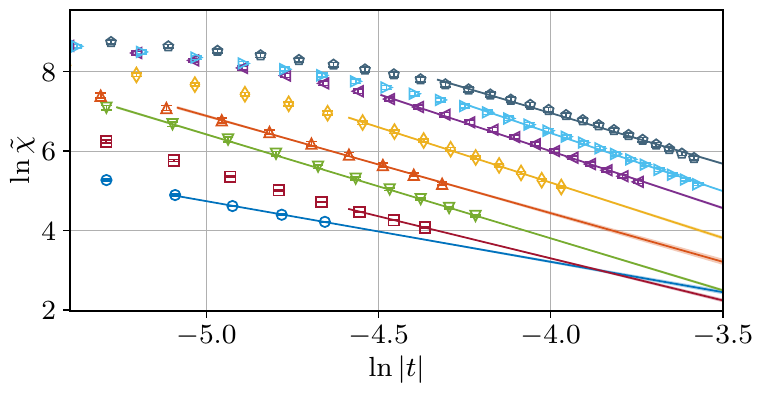}}
\vspace*{-3mm}
\subfloat{\includegraphics[scale=1]{figures/final/chiht_scaling/chiht_pd_for_legend}}
\caption
[Examples of the fits of $tilde \chi (t)$.]
{(Colour online) Examples of the fits of $[\tilde \chi (t)]$ with the ansatz 
(\ref{eq:temperature_scaling_analysis:chi_fit_ansatz}).
For each concentration of defects $p_d$ the minimum $|t|_{\rm min}$ for which
$\chi^2 \le 1.0$ was true for the first time is used in the plots.}
\label{fig:temperature_scaling_analysis:chiht_exp_fit_examples}
\end{figure}

\begin{figure}[!h]
\centering
\subfloat[$a = \infty$]{\includegraphics[scale=0.9]{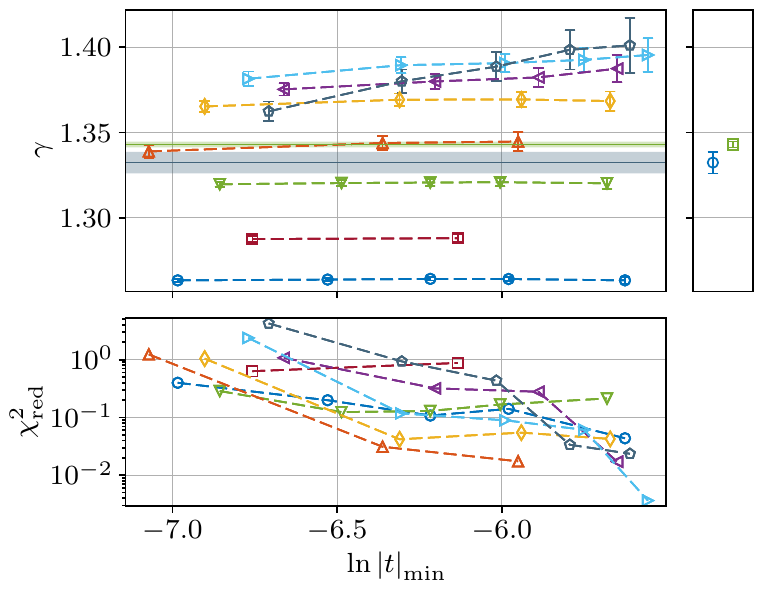}}
\subfloat[$a = 3.5$]{\includegraphics[scale=0.9]{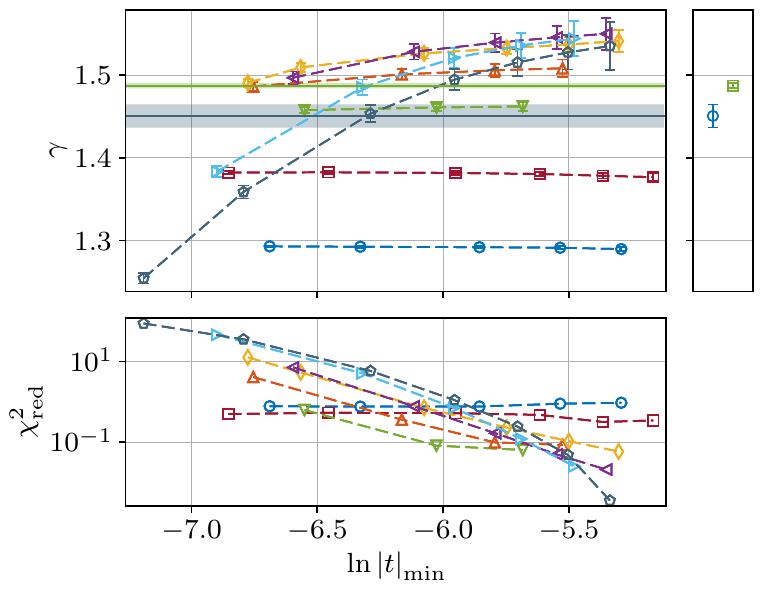}}
\vspace*{-3mm}
\subfloat[$a = 3.0$]{\includegraphics[scale=0.9]{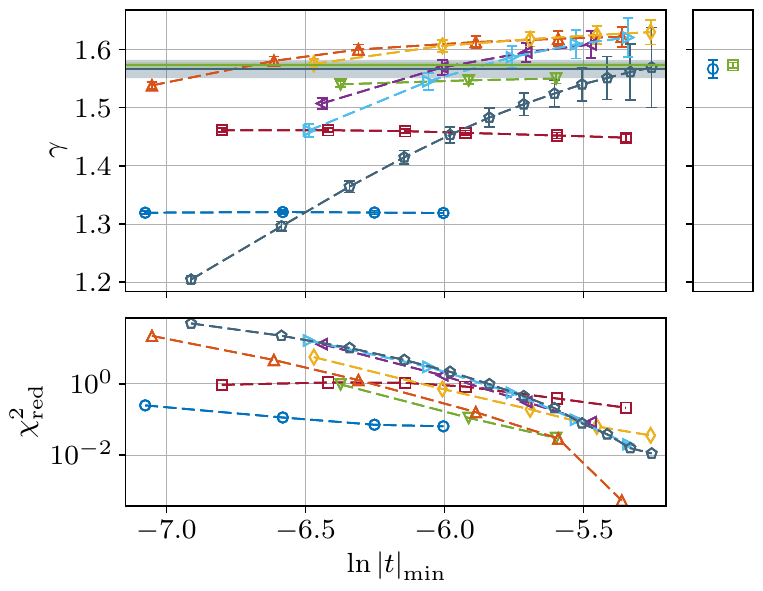}}
\subfloat[$a = 2.5$]{\includegraphics[scale=0.9]{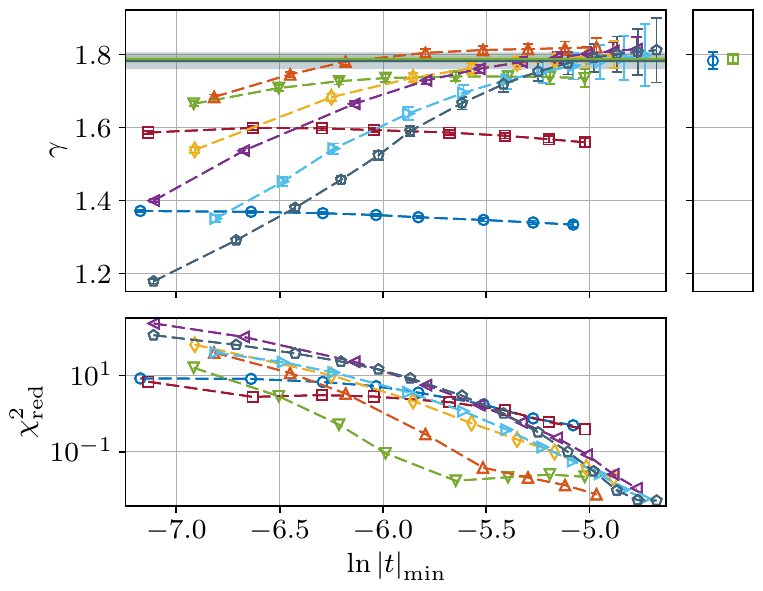}}
\vspace*{-3mm}
\subfloat[$a = 2.0$]{\includegraphics[scale=0.9]{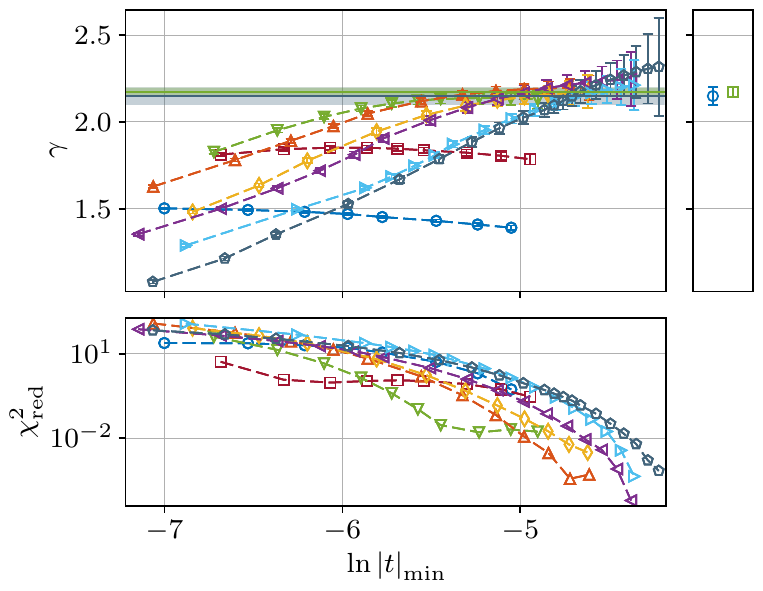}}
\subfloat[$a = 1.5$]{\includegraphics[scale=0.9]{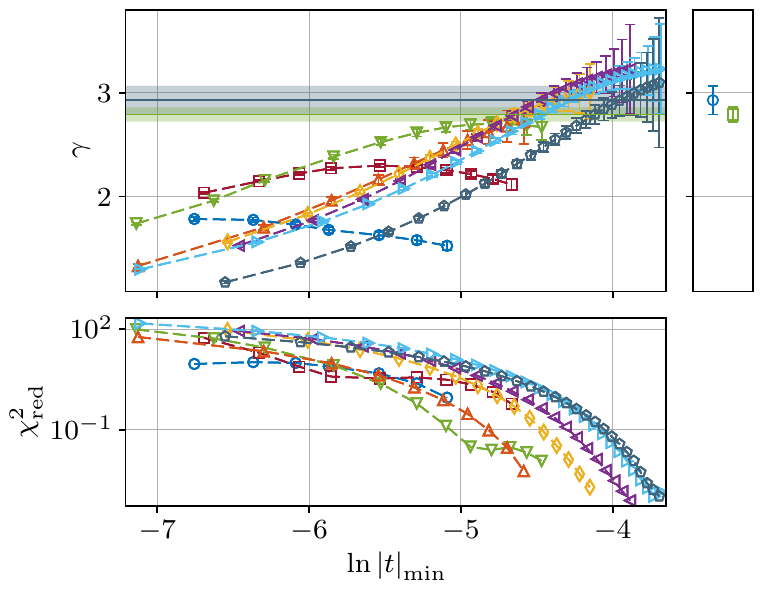}}
\vspace*{-3mm}
\subfloat{\includegraphics[scale=1]{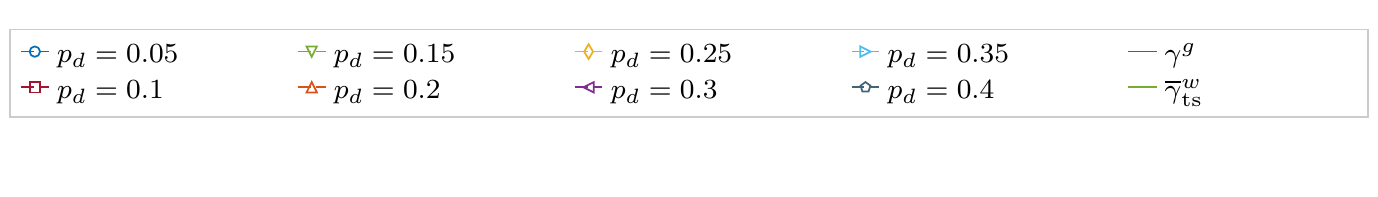}}
\vspace*{-5mm}
\caption[Final results of the fits of $\tilde \chi (t)$ for different $|t|_{\rm min}$.]
{(Colour online) Final results of the fits of $[\tilde \chi (t)]$ with the ansatz 
(\ref{eq:temperature_scaling_analysis:chi_fit_ansatz}) for different 
$|t|_{\rm min}$ and all defect concentrations $p_d$.
The weighted means $\overline{\gamma}^w_{\text{ts}}$ over all 
concentrations with $p_d \geqslant 0.15$ and for the largest possible 
$|t|_{\rm min}$ are shown together with the result from the FSS analysis 
$\gamma^{\rm g}$. The narrow right-hand panels show a separate comparison 
between the different estimates for $\gamma$ which are plotted as 
horizontal lines in the main plots.}
\label{fig:temperature_scaling_analysis:gamma_t_min_dependence}
\end{figure}

For the fits of the susceptibility $[\tilde \chi(T)]$ in the high-temperature 
phase, 
we used the ansatz
\begin{align}
	\ln [\tilde \chi(t)]  = B - \gamma \ln |t|
\label{eq:temperature_scaling_analysis:chi_fit_ansatz}
\end{align}
and again performed individual fits for all correlation exponents $a$ and
defect concentrations $p_d$. As in the case of $\xi(T)$, we varied the minimal
$|t|_{\rm min}$ included in the fits to see the asymptotic behaviour.
\clearpage 
\clearpage 
%
The susceptibility as function of $t$ is presented in 
figure \ref{fig:temperature_scaling_analysis:chiht_exp_t_curves}.
It has the same qualitative behaviour as $\xi$---for different $p_d$ the 
curves cross each other at $t \approx 0$. Note that as in the case of $\xi$, the 
definition of $\tilde \chi$ is valid only in the high-temperature phase, and 
we only extended the $T$ values below $T_c$ in order to see the crossing 
points better. In 
figure \ref{fig:temperature_scaling_analysis:chiht_exp_fit_examples} we show 
examples of the fits for all correlation exponents $a$ and defect
concentrations $p_d$. The estimates of the critical exponent $\gamma$ in 
dependence on the chosen $|t|_{\rm min}$ are presented in 
figure \ref{fig:temperature_scaling_analysis:gamma_t_min_dependence}.
The error weighted means of $\gamma$ over all $p_d \geqslant 0.15$ with the 
largest $|t|_{\rm min}$ are summarized in 
table \ref{tab:temperature_scaling_analysis:compare_fss_ts}.

The first observation is the same as in the case of $\xi$: The smallest 
concentrations $p_d \leqslant 0.1$ show a crossover behaviour and therefore we 
excluded them in the weighted mean. Again, the curves do not reach the 
asymptotic values even for the largest $|t|_{\rm min}$. The final weighted 
mean estimates $\overline{\gamma}^w_{\text{ts}}$ lie slightly above the 
corrected global fit estimates $\gamma^{\rm g}$ from the FSS analysis except 
for the case of $a = 1.5$. They match very well for the correlation 
exponents in the range $2.0 \leqslant a \leqslant 3.0$ but do not agree well in the 
uncorrelated case, $a = \infty$. The crossover region with 
$a \approx 3.0 -  3.5$ 
shows the largest deviations between $\overline{\gamma}^w_{\text{ts}}$ 
and $\gamma^{\rm g}$. In general, however, the qualitative cross-check 
with temperature scaling does support our estimates from the FSS 
analysis \cite{own_prb22}. Unfortunately, here we cannot compare with 
the individual uncorrected fit ansatz in the FSS case, since we have 
not performed it for the critical exponent $\gamma$.
\newpage

\begin{table}[!htb]
\centering
\small
\caption[Comparison of final estimates for the critical exponents $\nu$ and
$\gamma$ from temperature-scaling and FSS analyses.] 
{Comparison of final estimates for the critical exponents $\nu$ and 
$\gamma$ from temperature-scaling (labeled with ``ts'') and FSS analyses
for several (input) correlation exponents $a$ together with
their actual numerically measured values  $\overline{a}$.
The weighted means 
$\overline{\nu}^w_{\text{ts}}$ 
and
$\overline{\gamma}^w_{\text{ts}}$ 
over $p_d \ge 0.15$ 
were calculated over the estimates with 
the maximal $|t|_{\rm min}$ (and hence three degrees of freedom).
The FSS estimates $\nu^{\rm g}$ and $\gamma^{\rm g}$ 
are taken from  \cite{own_prb22} and
$\overline{\nu}^w_{\rm lin}$ and $\overline{\nu}^w$ from
 \cite{SK_Thesis}. 
}
\label{tab:temperature_scaling_analysis:compare_fss_ts}
%
%
\begin{tabular}{llllllll}
\hline\hline
\multicolumn{1}{c}{$a$} & 
\multicolumn{1}{c}{$\overline{a}$} & 
\multicolumn{1}{c}{$\overline{\nu}^w_{\text{ts}}$} & 
\multicolumn{1}{c}{$\overline{\nu}^w_{\rm lin}$} & 
\multicolumn{1}{c}{$\overline{\nu}^w$} & 
\multicolumn{1}{c}{$\nu^{\rm g}$} & 
\multicolumn{1}{c}{$\overline{\gamma}^w_{\text{ts}}$} & 
\multicolumn{1}{c}{$\gamma^{\rm g}$} \\
\hline
$\infty$ & $\infty$
     & 0.6928(17) & 0.6913(15) & 0.6843(31) & 0.6831(30)  & 1.3430(18) & 1.3324(64) \\
3.5 & 3.30(18)
     & 0.7557(25) & 0.7427(25) & 0.7122(49) & 0.7117(49)  & 1.4875(33) & 1.451(15)  \\
3.0 & 2.910(96)
     & 0.7898(34) & 0.7812(35) & 0.7532(53) & 0.7484(52)  & 1.5726(50) & 1.566(16)  \\
2.5 & 2.451(26)
     & 0.8905(82) & 0.8887(61) & 0.8725(96) & 0.8719(96)  & 1.787(11)  & 1.783(24)  \\
2.0 & 1.979(18)
     & 1.073(23)  & 1.079(14)  & 1.067(23)  & 1.060(23)   & 2.171(27)  & 2.149(51)  \\
1.5 & 1.500(30)
     & 1.348(61)  & 1.449(32)  & 1.435(56)  & 1.421(55)   & 2.791(70)  & 2.93(14)   \\
\hline\hline
\end{tabular}
\end{table}

%


\section{Conclusions}
\label{sec:concl}

The main merit of temperature
scaling is its conceptual simplicity. It provides direct estimates of the
critical exponents $\nu$, $\gamma$, \dots, whereas in the complementary FSS 
approach they
can only be computed from the fitted exponent ratios $1/\nu$,  
$\gamma/\nu$, \dots which requires some care with statistical error
propagation.
A drawback of temperature scaling is that the 
determination of a suitable fit interval requires care at both ends. 
If the included temperatures $T$ are too far away from $T_c$, 
corrections-to-scaling 
cannot be neglected. At the 
other end, the included $T$ values should not be too close to $T_c$,
because finite-size effects become important.
By contrast, in the FSS approach, only the lower end of the fit interval 
needs to be controlled: The minimal lattice size $L$ must be large enough to 
avoid sizeable corrections-to-scaling.
In this case,  in FSS analyses one  generically deals with simple linear 
two-parameter fits (since $T_c$ only enters indirectly), whereas in
temperature scaling, this is only possible when $T_c$ is known from other
sources (otherwise more cumbersome non-linear three-parameter fits are 
necessary).
Of course, in both approaches, the situation becomes 
more complicated when corrections-to-scaling should be included since 
this introduces additional parameters and generically requires a non-linear 
many-parameter fitting. 

Here, we successfully used the temperature-scaling analysis to validate our 
FSS
results \cite{own_prb22} for the critical exponents 
$\nu$ and $\gamma$. Considering that
mostly different data entered the analysis and also different observables were 
used, i.e., the correlation length $\xi$ that was not studied in the 
FSS approach and the high-temperature definition of the 
susceptibility, we can 
clearly solidify our 
FSS results. Additionally, the
critical temperatures estimated in the 
FSS study \cite{own_prb22} were confirmed to 
be reasonably accurate to be used in the temperature-scaling analysis.

The available data turned out, however, to
be not sufficiently accurate to perform fits including corrections-to-scaling,
and the uncorrected fits show a clear dependence on the  temperature range used. 
In order to improve the temperature-scaling analysis, we would need more
simulated temperatures and also change the setup of the entire simulation 
process by using more disorder realizations instead of longer measurement 
time series for each realization, which is planned for future work. 


\section*{Acknowledgements}
\label{sec:acknowl}

This paper is dedicated to Professor Bertrand Berche on the occasion of his 
60th birthday,
with whom the basics of the present work was laid 20 years
ago in a series of joint papers on uncorrelated bond disorder.
It has been always a pleasure to collaborate with Bertrand --- and to enjoy
the after-work activities!

We thank 
the Max Planck Society, 
the Max Planck Institute for Mathematics in the Sciences (MIS),
and especially the Graduate School IMPRS-MIS
for financial support of SK and providing the computational resources at 
the Max Planck Computing and Data Facility in Munich. 

We also gratefully acknowledge further support by the 
Deutsch-Franz\"osische Hochschule (DFH-UFA) through the Doctoral 
College ``$\mathbb{L}^4$'' under Grant No.\ CDFA-02-07 which is co-directed
by Bertrand. Many thanks go to him, Malte Henkel, Dragi Karevski,
and Christophe Chatelain for many fruitful discussions and hosting 
SK's visit in Nancy.
We also thank Yurko Holovatch, Viktoria Blavatska, and Mikhail Nalimov 
for interesting discussions that contributed to a deeper insight into 
the topic.



\ukrainianpart

\title[Степенева cкорельована невпорядкована 3D модель Ізінга]%
{Аналіз температурного скейлінгу тривимірної невпорядкованої моделі Ізінга зі степеневими скорельованими дефектами}
\author{С. Казмін\refaddr{label1,label2},
	В. Янке\refaddr{label1}}

\addresses{
\addr{label1} Інститут теоретичної фізики, Університет Лейпцига, IPF 231101,
04801 Лейпциг, Німеччина
\addr{label2} Німецький некомерційний науково-дослідний центр біологічних матеріалів, Торгауер штр. 116, 04347 Лейпциг, Німеччина
}

\makeukrtitle

\begin{abstract}
	Ми розглядаємо тривимірну розведену модель Ізінга зі степеневими кореляціями дефектів та досліджуємо критичну поведінку другого моменту кореляційної довжини і магнітної сприйнятливості у високотемпературній фазі.
	Порівнюючи отримані для різних інтенсивностей кореляції дефектів критичні показники $\nu$ та $\gamma$ з результатами нашого попереднього дослідження скінченно-вимірного скейлінгу, робимо узгоджені оцінки цих показників.
	\keywords тривимірна розведена модель Ізінга, далекосяжні кореляції, моделювання Монте-Карло, температурний скейлінг, критичні показники
\end{abstract}

\lastpage
\end{document}